\documentclass{article}
\PassOptionsToPackage{numbers, compress}{natbib}
\usepackage[dblblindworkshop, final]{neurips_2025}
\usepackage[utf8]{inputenc}
\usepackage[T1]{fontenc} 
\usepackage{hyperref}    
\usepackage{url}           
\usepackage{booktabs}       
\usepackage{amsfonts}    
\usepackage{nicefrac}    
\usepackage{microtype}     
\usepackage{multirow}
\usepackage{xcolor}
\usepackage{amsmath}
\usepackage{algorithm}
\usepackage{algpseudocode}
\usepackage{subcaption}
\usepackage{lineno}
\usepackage{subcaption}
\usepackage{graphicx}

\workshoptitle{Efficient Reasoning}
\title{UniFormer: Unified and Efficient Transformer for Reasoning Across General and Custom Computing}

\author{
Zhuoheng~Ran\thanks{Corresponding author}\\
Department of Electrical Engineering\\
City University of Hong Kong\\
\texttt{zhuoheran2-c@my.cityu.edu.hk}\\
\And
Chong~Wu\\
Department of Electrical Engineering\\
City University of Hong Kong\\
\texttt{chongwu2-c@my.cityu.edu.hk}\\
\And
Renjie~Xu\\
Department of Electrical Engineering\\
City University of Hong Kong\\
\texttt{harryxu950510@gmail.com}\\
\And
Maolin~Che\\
School of Mathematics and Statistics\\
Guizhou University\\
\texttt{mlche@gzu.edu.cn}\\
\AND
Hong~Yan\\
Department of Electrical Engineering\\
City University of Hong Kong\\
\texttt{h.yan@cityu.edu.hk}\\
}

\begin{document}
\maketitle

\begin{abstract}
The success of neural networks such as convolutional neural networks (CNNs) has been largely attributed to their effective and widespread deployment on customised computing platforms, including field-programmable gate arrays (FPGAs) and application-specific integrated circuits (ASICs). In the current era, Transformer-based architectures underpin the majority of state-of-the-art (SOTA) larger models that are also increasingly deployed on customised computing hardware for low-power and real-time applications. However, the fundamentally different parallel computation paradigms between general-purpose and customised computing often lead to compromises in model transfer and deployability, which typically come at the cost of complexity, efficiency or accuracy. Moreover, many cross-platform optimisation principles have also remained underexplored in existing studies. This paper introduces UniFormer, a unified and efficient Transformer architecture for both general-purpose and customised computing platforms. By enabling higher parallelism and compute–storage fusion, UniFormer achieves state-of-the-art (SOTA) accuracy and latency on GPUs while exhibiting strong adaptability on FPGAs. To the best of our knowledge, this paper is the first efficient Transformer work that jointly considers both general-purpose and customised computing architectures.
\end{abstract}

\section{Introduction}
The increasing usage of generative AI (GenAI) has posed significant challenges to both the cost for industry and the 2030 carbon neutrality goals set by governments. Fundamentally, this stems from the attention mechanism that endows Transformers with robust modelling capabilities and underpin state-of-the-art (SOTA) performance in large language models (LLMs)~\cite{huang2024mindmerger,du2024stacking,jiang2024llms}, Vision Transformers (ViTs)~\cite{diko2024revit,hatamizadeh2025mambavision,nie2024scopevit} and other applications~\cite{ganz2024question,xu2024utv,wu2020fuzzy,srivastava2024omnivec2,ran2022performance,waligora2024joint,anonymous2025gpstpca,xu2025lu,ran2022twoway,xu2025sparsegeohopca,lu2025rsvp,ran2023medical}. However, the attention mechanism also contributes as a major bottleneck and introduces significant computing complexity and memory overhead~\cite{papa2024survey,qin2024ayaka}.
\begin{figure}[t]
  \centering
  \begin{minipage}[c]{0.48\linewidth}
    \centering
    \includegraphics[width=\linewidth]{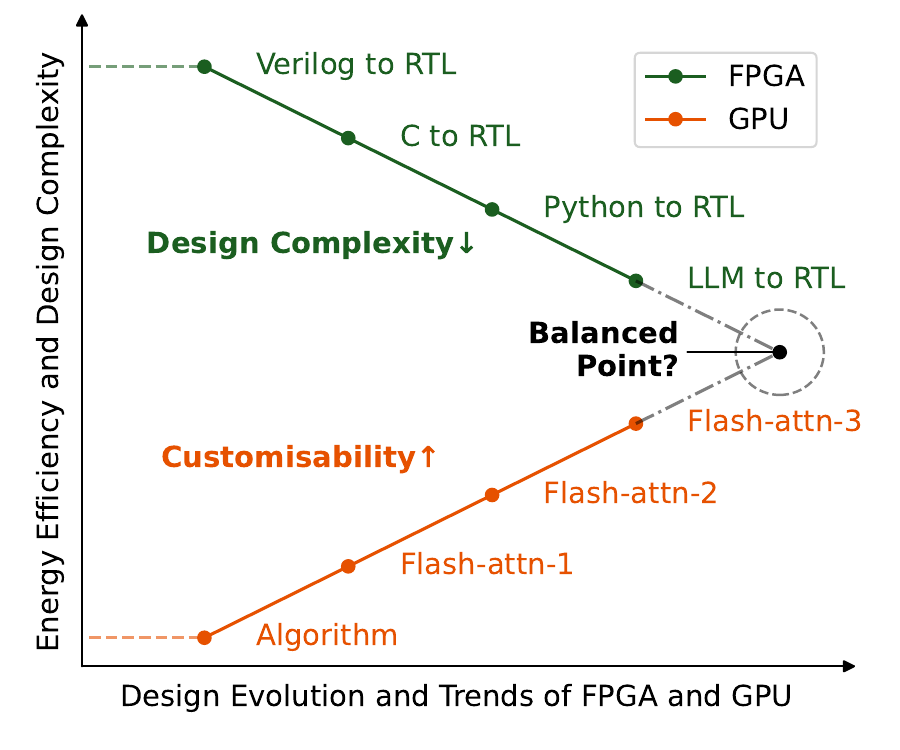}
  \end{minipage}
  \hfill
  \begin{minipage}[t]{0.51\linewidth}
    \centering
    \renewcommand{\arraystretch}{1.0}
    {\small
    \begin{tabular}{l l}
      \toprule
      Methods & Target Platforms \\
      \midrule
      DuSA~\cite{wudusa}                            & H20 GPU \\
      ELFATT~\cite{wu2025elfatt}                   & Jetson and  A100 GPU \\
      Keyformer~\cite{adnan2024keyformer}           & A100 GPU \\
      Nyströmformer~\cite{xiong2021nystromformer}   & GTX 1080Ti GPU \\
      CURSA~\cite{CURSA}                            & L20 and A800 GPU \\
      Swin Transformer~\cite{liu2022video}          & A100 GPU \\
      Longformer~\cite{zhu2021long}                 & RTX 8000 GPU \\
      FlatFormer~\cite{liu2023flatformer}           & Jetson and A6000 GPU \\
      Skyformer~\cite{chen2021skyformer}            & V100 GPU \\
      Polynormer~\cite{dengpolynormer}              & A6000 GPU and CPU \\
      EfficientViT~\cite{cai2023efficientvit}       & Jetson GPU and CPU \\
      \textbf{UniFormer (Ours)}                     & \textbf{GPU / FPGA} \\
      \bottomrule
    \end{tabular}
    }
  \end{minipage}
  \caption{\textbf{Problem setting (left):} Recent efficient Transformer works leverage customised GPU kernels to maximise parallelism and memory utilisation, while FPGA designs can already enable fine-grained dataflows and pipelined execution by high-level synthesis under significantly reduced computational complexity. Moreover, both general and customised computing architectures seek more advanced compute–storage fusion mechanisms to alleviate the common bandwidth bottleneck. 
\textbf{Existing works (right):} 
Representative efficient Transformer implementations with target platforms.}
  \label{fig:fpga_gpu_design_trend_table}
\end{figure}

During a long period prior to recent advances, one of the key factors behind the success of many state-of-the-art (SOTA) neural networks, such as convolutional neural network (CNN), was their effective deployment on customised computing hardware that included field-programmable gate arrays (FPGAs)~\cite{bosio2024nn2fpga,chen2023graph,montgomerie2023satay} and application-specific integrated circuits (ASICs)~\cite{esmaeilzadeh2024performance,niknia2024asic,abubakar2022746} which effectively alleviated the challenges faced by neural networks in energy efficiency and cost effectiveness. In contrast, the vanilla Transformer and existing efficient Transformers are poorly considered and adapted for customised hardware due to structural incompatibilities in distinct computational paradigms~\cite{yang2024glitches,bhowmick2023optimizing}.

Early work~\cite{li2022auto,wang2022via,marino2023me} on customised hardware typically sought to rearrange or replace certain functions to align with the vanilla Transformer, but this often incurred substantial design complexity. 
Although recent efficient Transformer variants have introduced various approximations to reduce computational complexity, many of them remain poorly supported on customised hardware~\cite{zhang2024efficientvit,ran2024ro,xu2024cur,ran2022hardware} 
due to the issues in expensive matrix inversions, limited data reuse and excessive data dependencies. By incorporating kernel-based attention functions (e.g., ReLU~\cite{shen2023study,zhang2025inhibidistilbert,wortsman2023replacing}) and hardware-friendly operations such as convolutions~\cite{khan2023survey,son2024csta,yuan2023effective}, recent efficient Transformer works have achieved state-of-the-art (SOTA) performance on GPUs and demonstrated potential for customised hardware deployment~\cite{wu2025elfatt,pwu2025elfatt,wu2025elfattHK,shao2024fpga,ran2025fastvit}. However, these variants were primarily designed for GPUs, and their efficiency gains often come at the cost of accuracy. This trade-off becomes particularly pronounced when low-bit quantisation is required for deployment on resource-constrained reconfigurable platforms on customised hardware.

Furthermore, recent advances in efficient Transformers and hardware accelerators also reveal converging optimisation principles across general-purpose and customised computing architectures, as illustrated in Figure~\ref{fig:fpga_gpu_design_trend_table} (left). Motivated by these observations and based on the architecture of ELFATT~\cite{wu2025elfatt}, this work makes the following contributions: 
a) We propose \textbf{UniFormer}, a dual-branch attention mechanism that adopts matrix multiplication as the fundamental optimisation primitive to minimise computational switching. The fusion of global linear attention and local block-wise attention reduces data dependency and enhances parallelism to enable efficient compute--storage fusion across diverse workloads.
b) For GPU deployments, we develop a Triton-based GPU kernel for the global attention branch, which achieves significantly lower latency than the PyTorch implementation. Across various sequence lengths and batch configurations, the proposed kernel delivers a throughput improvement ranging from \textbf{1.79$\times$} to \textbf{2.36$\times$}, while maintaining comparable or higher Top-1 accuracy on ImageNet.
c) For FPGA deployments, UniFormer achieves \textbf{470$\times$} acceleration and energy efficiency gains over vanilla Transformers with cross-platform deployment capabilities.

To the best of our knowledge, this is the \textbf{first} efficient Transformer work that takes into account both general-purpose and customised computing architectures.

\section{Preliminaries: Matrix Multiplication in Attention and Parallel Computing}
\begin{figure}[t]
\centering
\includegraphics[width=\linewidth]{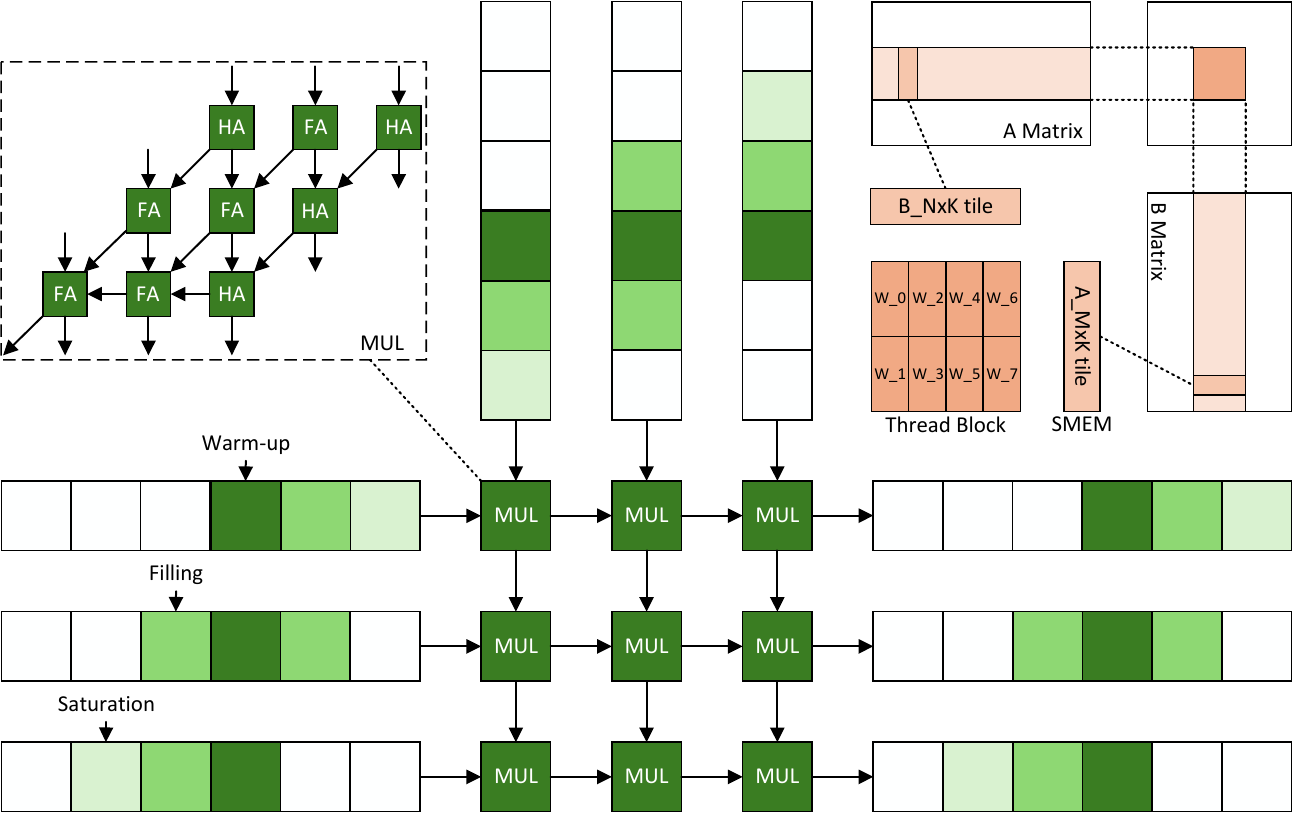}
\caption{Matrix multiplication on parallel computing architectures, such as FPGAs and GPUs, can be further decomposed into fine-grained parallel blocks and multiply-add operations to enable customised and hardware-aware optimisations. Favourable trade-offs between design complexity and efficiency make matrix multiplication an efficient computational primitive across both architectures.}
\label{fig:matrix_multiplication}
\end{figure}
Many operations introduced in recent efficient Transformer variants are not directly applicable to customised computing platforms, as they often rely on some improved functions that are poorly supported or inefficient on customised computing platform. Among the solutions and operations that remain feasible, hybrid structures such as ReLU-based attention mechanisms and convolutional operations show promising potential as hardware-friendly alternatives. However, these designs still incur non-trivial overhead due to frequent switching between varying types of operations and may suffer from degradation in both accuracy and generalisation as a result of deviations from the original GEMM-based similarity computation. These limitations are further exacerbated on customised hardware, where fixed-point arithmetic and quantisation are indispensable for efficient deployment.

In contrast, GEMM can serve as the "highest common factor" for optimisation in general-purpose computing architectures and customised computing architectures for the following reasons: First, since hybrid designs with diverse operators often lead to computational imbalance and irregular dataflows, incurring additional scheduling and communication overhead, GEMM preserves the mathematical fidelity of attention and minimises accuracy loss compared to replacing it with alternative operators such as convolutions or other operations. Second, many challenges common to GPUs and FPGAs, such as limited off-chip bandwidth and the necessity of compute–storage fusion, can be further addressed through well-optimised GEMM computing components. Moreover, GEMM has been extensively studied and highly optimised for both GPUs and FPGAs, making it a natural optimisation target. On GPUs, tensor cores enable cycle-level fused multiply–accumulate operations with warp-level collaboration, shared memory tiling and WMMA APIs to support massively parallel GEMM execution. Whereas in FPGAs, GEMM can be mapped to systolic arrays that are deeply pipelined to maximise throughput and energy efficiency, with each multiplication unit further decomposed into combinational adders to exploit fine-grained parallelism. In addition, GEMM has been a central focus in approximate computing and low-precision arithmetic research, where units with reduced bitwidth multiply-accumulate maintain efficiency without compromising accuracy. As illustrated in Figure~\ref{fig:matrix_multiplication}, these facts and paradigms suggest that GEMM can serve as a minimal optimised primitive and is well-suited for efficient Transformers deployed across GPUs and FPGAs.

\begin{figure}[t]
    \centering
    \includegraphics[width=\linewidth]{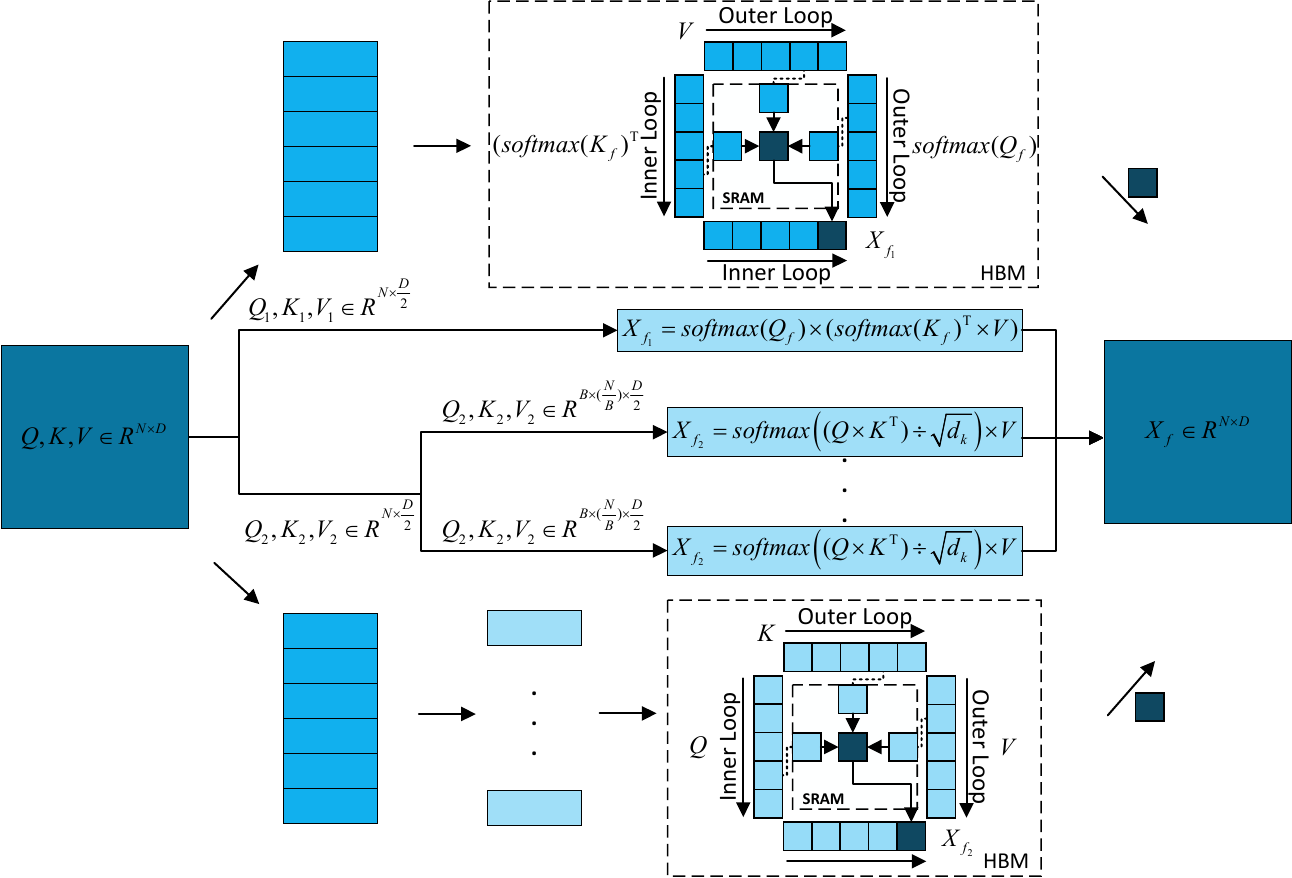}
    \caption{The proposed dual-branch Transformers. The input features are split into more parallel streams and can be fused with Triton kernels with inner-loop and outer-loop scheduling strategies to maximise data and memory usage (e.g., SRAM and HBM in GPUs, BRAM and DRAM in FPGAs).}
    \label{fig:framework}
\end{figure}

\section{Methodology}
We discuss in Preliminaries with Figure~\ref{fig:matrix_multiplication} the role of general matrix multiplication (GEMM) as the fundamental optimisation primitive across general and customised computing architectures. To achieve higher parallelism and compute-storage fusion, we propose a dual-branch attention mechanism that integrates local structure modelling with global dependency modelling based on matrix multiplication, which can be divided into local and global branches and is shown in Figure~\ref{fig:framework}. This solution can effectively mitigate full-size matrix multiplications and softmax operations in vanilla Transformers and significantly enhance computational fusion and parallelism. Specifically, the input sequence is bifurcated into two branches: the global branch captures long-range token dependencies using a linear complexity attention mechanism, while the local branch focuses on fine-grained context modelling through a block-wise strategy that partitions the sequence into parallel and independent blocks. 
This hierarchical decomposition facilitates extensive parallel computation and preserves global dependencies without sacrificing local precision. For improved fusion of compute and storage, we implement attention computations within the global branch using a Triton-based fused kernel inspired by FlashAttention, maximising on-chip data reuse and minimising memory transfer overhead on both GPU and FPGA targets. The overall architecture of UniFormer is illustrated in Figure~\ref{fig:framework}.

Given $Q,K,V \in \mathbb{R}^{B \times N \times d_k}$, we partition the sequence into $T$ windows of length $N_w = s^2$:
\begin{equation}
(Q_{b\_all}, K_{b\_all}, V_{b\_all}) = \mathrm{blockify}(Q, K, V)
\in \mathbb{R}^{(B \cdot T) \times N_w \times d_k}.
\label{eq:blockify_unified_main}
\end{equation}

\paragraph{Local branch.}
Within each window $i$, we apply scaled dot-product attention:
\begin{equation}
X_{b}^{(i)} =
\mathrm{softmax}\!\left(\frac{Q_b^{(i)} (K_b^{(i)})^{\!\top}}{\sqrt{d_k}}\right)
V_b^{(i)}.
\label{eq:block_attention_comp_unified_main}
\end{equation}
The window outputs are merged to the full sequence by deblocking.

\begin{algorithm}[htbp]
\caption{Algorithm 1: Block-Local Attention}
\label{alg:local_block_attention}
\begin{algorithmic}[1]
\Statex \textbf{Input:} $Q,K,V \in \mathbb{R}^{B \times N \times d_k}$; block size $\text{BLOCK\_SIZE}$ (and tiling $\text{BLOCK\_N}$)  
\Statex \textbf{Output:} Local-branch output $X_1 \in \mathbb{R}^{B \times N \times d_k}$
\State $(Q_{\text{loc}},K_{\text{loc}},V_{\text{loc}}) \gets \text{SelectPortionForLocal}(Q,K,V)$
\State $(Q_b,K_b,V_b) \gets \text{blockify}(Q_{\text{loc}},K_{\text{loc}},V_{\text{loc}})$
\State Initialize $X_{1,b}$ with the shape of $Q_b$; \quad $s \gets \sqrt{d_k}$
\For{\textbf{each} block index $i$ \textbf{in parallel}}
  \State $Q_i \gets Q_b[i,:,:],\; K_i \gets K_b[i,:,:],\; V_i \gets V_b[i,:,:]$
  \For{\textbf{each} row $q_r$ of $Q_i$}
    \State $m \gets -\infty$;\; $l \gets 0$;\; $\mathbf{acc} \gets \mathbf{0} \in \mathbb{R}^{d_k}$
    \For{\textbf{each} sub-block $(K_{i,j},V_{i,j})$ tiled by $\text{BLOCK\_N}$}
      \State $S \gets q_r K_{i,j}^\top / s$;\; $m_{\text{old}} \gets m$;\; $m \gets \max(m, \max(S))$
      \State $P \gets \exp(S - m)$
      \State $\mathbf{acc} \gets \mathbf{acc}\cdot \exp(m_{\text{old}}-m) + P V_{i,j}$
      \State $l \gets l \cdot \exp(m_{\text{old}}-m) + \sum P$
    \EndFor
    \State $X_{1,b}[i,r,:] \gets \mathbf{acc}/l$
  \EndFor
\EndFor
\State $X_1 \gets \text{deblockify}(X_{1,b})$
\end{algorithmic}
\end{algorithm}

\paragraph{Global branch.}
In parallel, the global branch as linear attention~\cite{shen2021efficient,CURSA} on
$Q_g, K_g, V_g \in \mathbb{R}^{N_g \times d_k}$:
\begin{equation}
X_g
= \mathrm{softmax}_{\textsf{feat}}(Q_g)\,
\Big(\mathrm{softmax}_{\textsf{seq}}(K_g)^{\!\top} V_g\Big)
= \mathrm{softmax}_{\textsf{feat}}(Q_g)\, C_g,
\label{eq:global_linear_attention_factorized_unified_main}
\end{equation}
where $\mathrm{softmax}_{\textsf{feat}}$ normalises along the characteristic dimension ($d_k$),
$\mathrm{softmax}_{\textsf{seq}}$ normalises along the sequence dimension ($N_g$) and
$C_g=\mathrm{softmax}_{\textsf{seq}}(K_g)^{\!\top}V_g \in \mathbb{R}^{d_k \times d_k}$ is a global content matrix.

On GPUs, each of the $B \cdot T$ blocks can be mapped to independent thread blocks, and fused kernels can be designed to jointly execute local and global branches for higher throughput. In FPGAs, block computations are mapped to dedicated, deeply pipelined processing elements, enabling massive spatial parallelism. The reduced block size $N_w$ allows data to fit into fast on-chip memory (SRAM on GPUs, BRAM on FPGAs), alleviating HBM/DRAM bottlenecks on GPU/FPGA to reduce latency.

\begin{algorithm}[htbp]
\caption{Algorithm 2: Global Linear Attention}
\label{alg:global_linear_attention}
\begin{algorithmic}[1]
\Statex \textbf{Input:} $Q,K,V \in \mathbb{R}^{B \times N \times d_k}$  
\Statex \textbf{Output:} Global-branch output $X_2 \in \mathbb{R}^{B \times N \times d_k}$
\State $(Q_g,K_g,V_g) \gets \text{SelectPortionForGlobal}(Q,K,V)$
\State $Q_g' \gets \text{softmax}_{\text{feat}}(Q_g)$
\State Initialize $C_g \gets \mathbf{0} \in \mathbb{R}^{B \times d_k \times d_k}$
\For{\textbf{each} batch $b$ \textbf{in parallel}}
  \For{\textbf{each} feature $f=1\dots d_k$ \textbf{in parallel}}
    \State $m \gets -\infty$;\; $L' \gets 0$;\; $\mathbf{cv} \gets \mathbf{0} \in \mathbb{R}^{d_k}$
    \For{\textbf{each} sequence block $j$ of $(K_g[b],V_g[b])$}
      \State $K_{B,f} \gets$ $f$-th column of block $j$ in $K_g[b]$;\; $V_B \gets$ block $j$ in $V_g[b]$
      \State $m_{\text{old}} \gets m$;\; $m \gets \max(m, \max(K_{B,f}))$
      \State $P \gets \exp(K_{B,f}-m)$
      \State $\mathbf{cv} \gets \mathbf{cv}\cdot \exp(m_{\text{old}}-m) + P^\top V_B$
      \State $L' \gets L'\cdot \exp(m_{\text{old}}-m) + \sum P$
    \EndFor
    \State $C_g[b,f,:] \gets \mathbf{cv}/L'$
  \EndFor
\EndFor
\State $X_2 \gets Q_g' \, C_g$
\end{algorithmic}
\end{algorithm}

\section{Results}
\begin{figure}[t]
  \centering
  \includegraphics[width=0.75\linewidth]{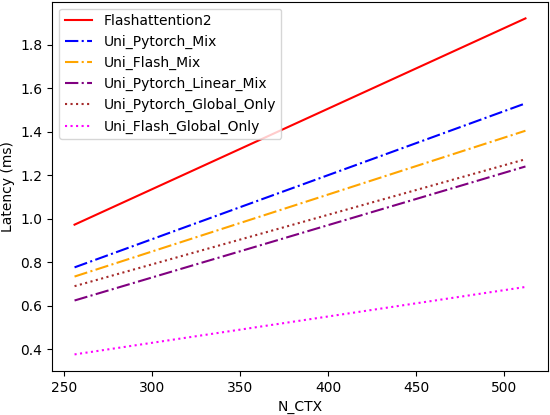}
  \caption{
    Throughput Analysis. Throughput performance on an H800 GPU (batch size = 64, head = 16, dimension = 64) under varying sequence lengths. 
    The plot illustrates the latency and throughput characteristics across model configurations. The more results are provided in Appendices~A.1 \& A.2.}
  \label{fig:h800_results}
\end{figure}

Based on a dual-branch attention mechanism, this paper proposes a Triton-based accelerated kernel for the global branch and the native compatibility of the local branch with FlashAttention2, which allows for flexible implementation as the accelerated kernel-based, Pytorch-based local or global attention without accuracy compromise. Figure~\ref{fig:h800_results} shows the various latency comparisons of these combinations on the H800 GPU (Batch Size = 64, Head = 16, Dim = 64) by evaluating different sequence lengths. Additional results on more GPUs are also available in Appendices~A.1 and ~A.2.

\textbf{Finding 1:} \textbf{Uni\_Flash\_Global\_Only} (Using only global attention with the global attention-accelerated kernel) achieves the lowest latency, which is consistent with the expectations and demonstrates the effectiveness of our linear attention Triton kernel design. \textbf{Uni\_Flash\_Mix} (Using global and local attention with the global and local attention-accelerated kernel) combines FlashAttention2 for the local branch with our Triton linear attention kernel design for the global branch, which outperforms official FlashAttention2 and standard Pytorch implementation \textbf{Uni\_Pytorch\_Mix} (Global and local attention by Pytorch) and again validates the effectiveness of our Triton kernel design. 

\textbf{Finding 2:} It is worth noting that \textbf{Uni\_Flash\_Mix}, which applies both local and global custom acceleration kernels but exhibits higher latency than \textbf{Uni\_Pytorch\_Linear\_Mix} (global and local attention with the global accelerated kernel and local Pytorch implementations). This suggests that further customised kernel designs do not always lead to acceleration gains, which we attribute to potential kernel contention and the highly optimised PyTorch's native attention operators. Another supporting proof is that \textbf{Uni\_Pytorch\_Linear\_Mix} achieves even lower latency than \textbf{Uni\_Pytorch\_Global\_Only} (which applies linear attention to both branches and serves as a control group for the latency test). This further indicates that combining PyTorch's highly optimised local attention with a customised global kernel yields a more efficient hybrid design that better exploits the structural characteristics of the UniFormer architecture. This observation also further reveals the potential downsides of overusing compute-storage fusion, such as kernel contention, thread-level interference and scheduling overhead.

\textbf{Finding 3:} Against the strong baseline \textbf{Local-Pytorch+Global-Pytorch} that uses PyTorch implementations for both branches, \textbf{Local-PyTorch+Global-Kernel} achieves a $1.37\times$ acceleration (4280 vs. 3119 img/s) with identical accuracy and model size. These results effectively demonstrate that combining the Triton-accelerated global attention kernel with an optimised local attention path is a superior design strategy to fully exploit GPU bandwidth without sacrificing model performance.

\begin{table*}[htbp]
  \centering
  \renewcommand{\arraystretch}{1.5}
  \setlength{\tabcolsep}{22pt}
  {\caption{Accuracy and throughput comparison.
    Accuracy and throughput results of representative state-of-the-art (SOTA) Transformer models on ImageNet 
    (224$\times$224 resolution, H20 GPU, batch size = 512). 
    All models have 20M--21M parameters for fair comparison.
  }
  \label{tab:h800_accuracy}
  \begin{tabular}{lcc}
    \toprule
    \textbf{Work} & \textbf{Acc. (\%)} & \textbf{FPS} \\
    \midrule
    EfficientViT~\cite{cai2023efficientvit}           & 82.0 & 1812 \\
    Agent Attention~\cite{han2024agent}               & 82.5 & 2395 \\
    Vanilla Transformer~\cite{vaswani2017attention}   & 82.9 & 2102 \\
    Flatten Transformer~\cite{han2023flatten}         & 82.8 & 1988 \\
    \midrule
    Local-Pytorch+Global-Pytorch                             & 82.9 & 3119 (All: 14.85s) \\
    \textbf{Local-PyTorch+Global-Kernel}                     & \textbf{82.9} & \textbf{4280 (All: 10.66s)} \\
    \bottomrule
  \end{tabular}
  }

\end{table*}

\textbf{Finding 4:} Table~\ref{tab:h800_accuracy} summarises the accuracy, throughput and model size of several state-of-the-art (SOTA) Transformer models on the ImageNet classification task, evaluated at input resolution 224 and batch size 512 on the H20 GPU. Compared with representative efficient Transformer baselines, including EfficientViT~\cite{cai2023efficientvit}, 
Agent Attention~\cite{han2024agent}, 
Vanilla Transformer~\cite{vaswani2017attention} and Flatten Transformer~\cite{han2023flatten}, 
our proposed \textbf{Local-Pytorch+Linear-Kernel} configuration achieves 
a throughput improvement ranging from $1.79\times$ to $2.36\times$,
while maintaining comparable or higher Top-1 accuracy.

\begin{table*}[htbp]
  \caption{Latency comparison of vanilla attention~\cite{vaswani2017attention} with ours.}
  \label{fig:fpga_latency_comparison}
  \centering
  \renewcommand{\arraystretch}{1.5}
  \setlength{\tabcolsep}{14pt}
  \begin{tabular}{c|cc|cc}
    \toprule
    \multirow{2}{*}{\textbf{Input Size}} & \multicolumn{2}{c|}{\textbf{Vanilla}} & \multicolumn{2}{c}{\textbf{UniFormer}} \\
     & \textbf{Cycles} & \textbf{Latency (ns)} & \textbf{Cycles} & \textbf{Latency (ns)} \\
    \midrule
    8    & 6{,}063     & $3.032 \times 10^{4}$   & 3{,}608      & $1.804 \times 10^{4}$ \\
    16   & 21{,}032    & $1.052 \times 10^{5}$   & 4{,}918      & $2.459 \times 10^{4}$ \\
    64   & 299{,}000   & $1.495 \times 10^{6}$   & 12{,}123     & $6.062 \times 10^{4}$ \\
    128  & 1{,}171{,}384 & $5.857 \times 10^{6}$ & 23{,}947     & $1.197 \times 10^{5}$ \\
    256  & 4{,}636{,}472 & $2.318 \times 10^{7}$ & 47{,}595     & $2.380 \times 10^{5}$ \\
    512  & 23{,}002{,}669 & $1.150 \times 10^{8}$ & 94{,}891     & $4.745 \times 10^{5}$ \\
    1024 & 89{,}259{,}053 & $4.463 \times 10^{8}$ & 189{,}483    & $9.474 \times 10^{5}$ \\
    \bottomrule
  \end{tabular}
\end{table*}
\textbf{Finding 5:}
As shown in Table~\ref{fig:fpga_latency_comparison} and Figure~\ref{fig:efficiency_comparison} in Appendix~A.3, the FPGA implementation reveals that the latency of vanilla attention grows quadratically with input size due to its complexity $O(N^2)$. 
In contrast, UniFormer maintains a nearly linear growth pattern with input size and achieves a speedup ranging from $180\times$ to $470\times$ and improved energy efficiency compared to vanilla attention. 

\section{Discussion}
This work presented UniFormer, a unified and efficient Transformer architecture that bridges the gap between general-purpose and customised computing platforms. By leveraging a dual-branch attention mechanism that integrates global linear attention with block-wise local feature extraction, UniFormer achieves high parallelism, reduced data dependency and effective compute-storage fusion without compromising accuracy. On GPUs, the proposed Triton-based fused kernels deliver substantial latency reduction and throughput improvement, while on FPGAs, UniFormer demonstrates notable acceleration and gains in energy efficiency under the Vitis HLS implementation. These results collectively validate the scalability, portability and cross-platform adaptability of the proposed architecture across distinct computing architecture. To the best of our knowledge, this is the first efficient Transformer work that takes into account both general-purpose and customised architectures.

\begin{ack}
This work is supported by Hong Kong Innovation and Technology Commission (InnoHK Project CIMDA), Hong Kong Research Grants Council (Project 11204821), and City University of Hong Kong (Projects 9610034 and 9610460).
\end{ack}

\bibliographystyle{unsrtnat}
\bibliography{references}

\appendix

\newpage
\section{Experiments Supplementary}
\subsection{Results on 4070Ti Super GPU}
\begin{figure*}[htbp]
    \centering

    \begin{subfigure}[t]{0.49\linewidth}
        \centering
        \includegraphics[width=\linewidth]{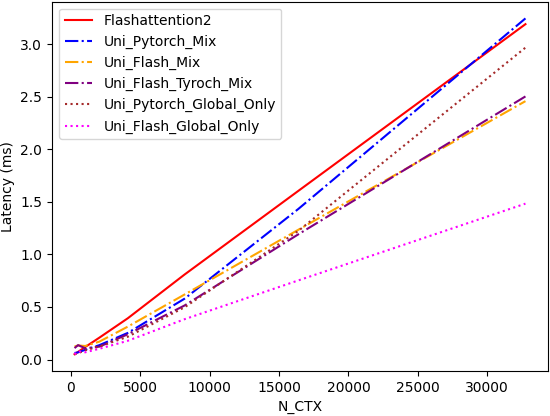}
        \caption{Batch Size-1, Head-16, Dimension-64}
    \end{subfigure}
    \hfill
    \begin{subfigure}[t]{0.49\linewidth}
        \centering
        \includegraphics[width=\linewidth]{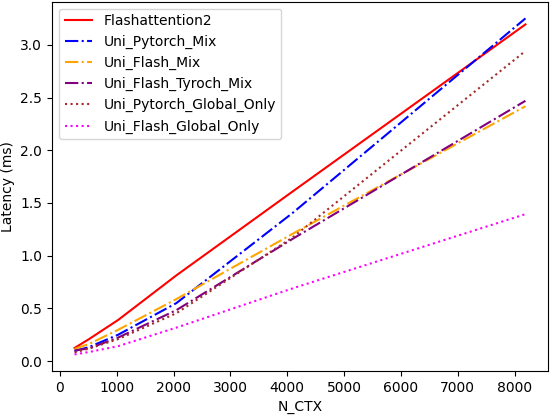}
        \caption{Batch Size-4, Head-16, Dimension-64}
    \end{subfigure}

    \vspace{1em}

    \begin{subfigure}[t]{0.49\linewidth}
        \centering
        \includegraphics[width=\linewidth]{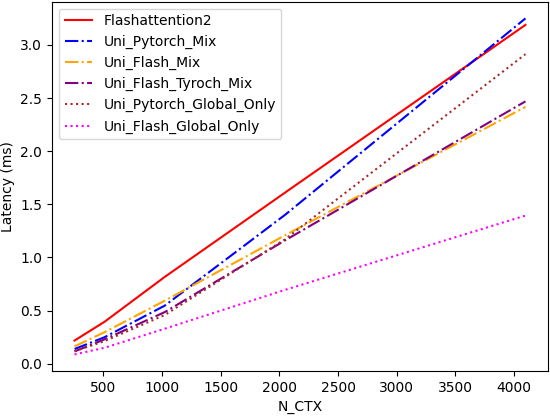}
        \caption{Batch Size-8, Head-16, Dimension-64}
    \end{subfigure}
    \hfill
    \begin{subfigure}[t]{0.49\linewidth}
        \centering
        \includegraphics[width=\linewidth]{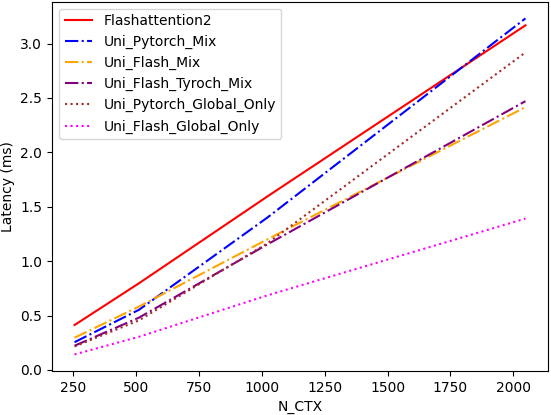}
        \caption{Batch Size-16, Head-16, Dimension-64}
    \end{subfigure}

    \vspace{1em}

    \begin{subfigure}[t]{0.49\linewidth}
        \centering
        \includegraphics[width=\linewidth]{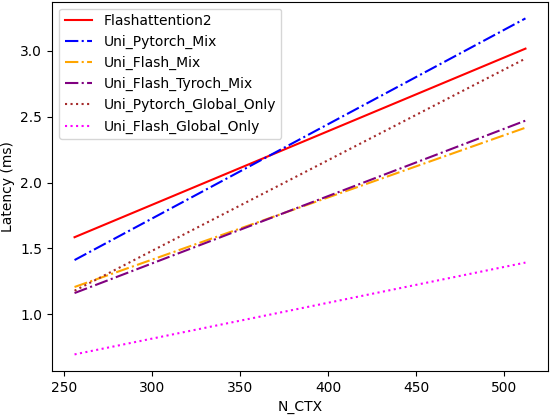}
        \caption{Batch Size-64, Head-16, Dimension-64}
    \end{subfigure}

    \caption{Latency vs. input context length ($N_{\text{CTX}}$) under five different attention on 4070 Ti Super.}
    \label{fig:latency_five_grid1}
\end{figure*}

\newpage
\subsection{Results on H800 GPU}
\begin{figure*}[htbp]
    \centering
    
    \begin{subfigure}[t]{0.49\linewidth}
        \centering
        \includegraphics[width=\linewidth]{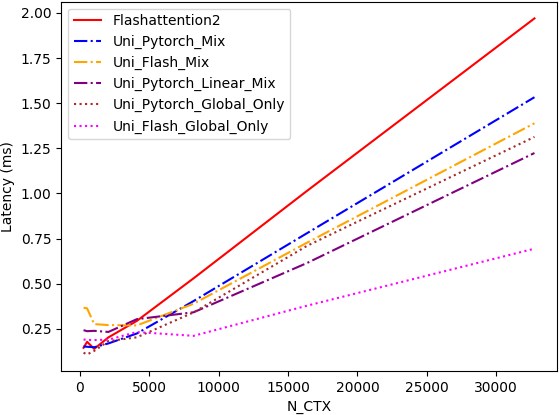}
        \caption{Batch Size-1, Head-16, Dimension-64}
    \end{subfigure}
    \hfill
    \begin{subfigure}[t]{0.49\linewidth}
        \centering
        \includegraphics[width=\linewidth]{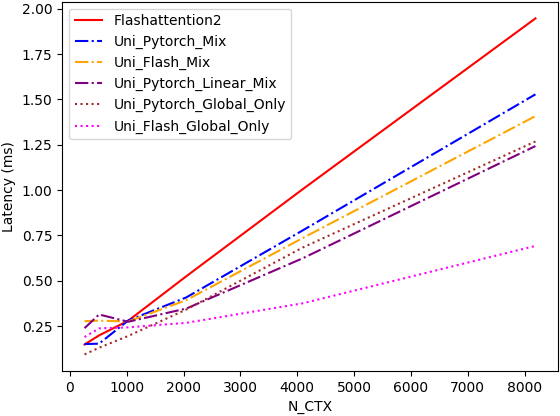}
        \caption{Batch Size-4, Head-16, Dimension-64}
    \end{subfigure}

    \vspace{1em}

    \begin{subfigure}[t]{0.49\linewidth}
        \centering
        \includegraphics[width=\linewidth]{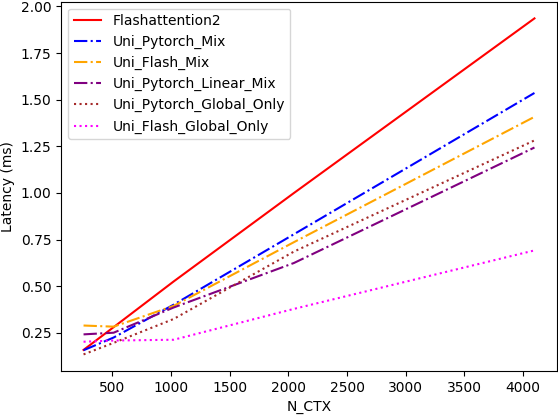}
        \caption{Batch Size-8, Head-16, Dimension-64}
    \end{subfigure}
    \hfill
    \begin{subfigure}[t]{0.49\linewidth}
        \centering
        \includegraphics[width=\linewidth]{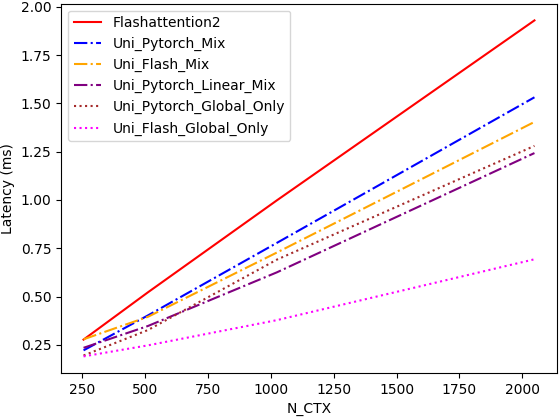}
        \caption{Batch Size-16, Head-16, Dimension-64}
    \end{subfigure}

    \vspace{1em}

    \begin{subfigure}[t]{0.49\linewidth}
        \centering
        \includegraphics[width=\linewidth]{H800-64-500-16-64.png}
        \caption{Batch Size-64, Head-16, Dimension-64}
    \end{subfigure}

    \caption{Latency vs. input context length ($N_{\text{CTX}}$) under five different attention on H800 GPU.}
    \label{fig:latency_five_grid2}
\end{figure*}

\newpage
\subsection{Additional Experimental Results in FPGA implementation}
\begin{figure*}[htbp]
    \centering
    \includegraphics[width=1\linewidth]{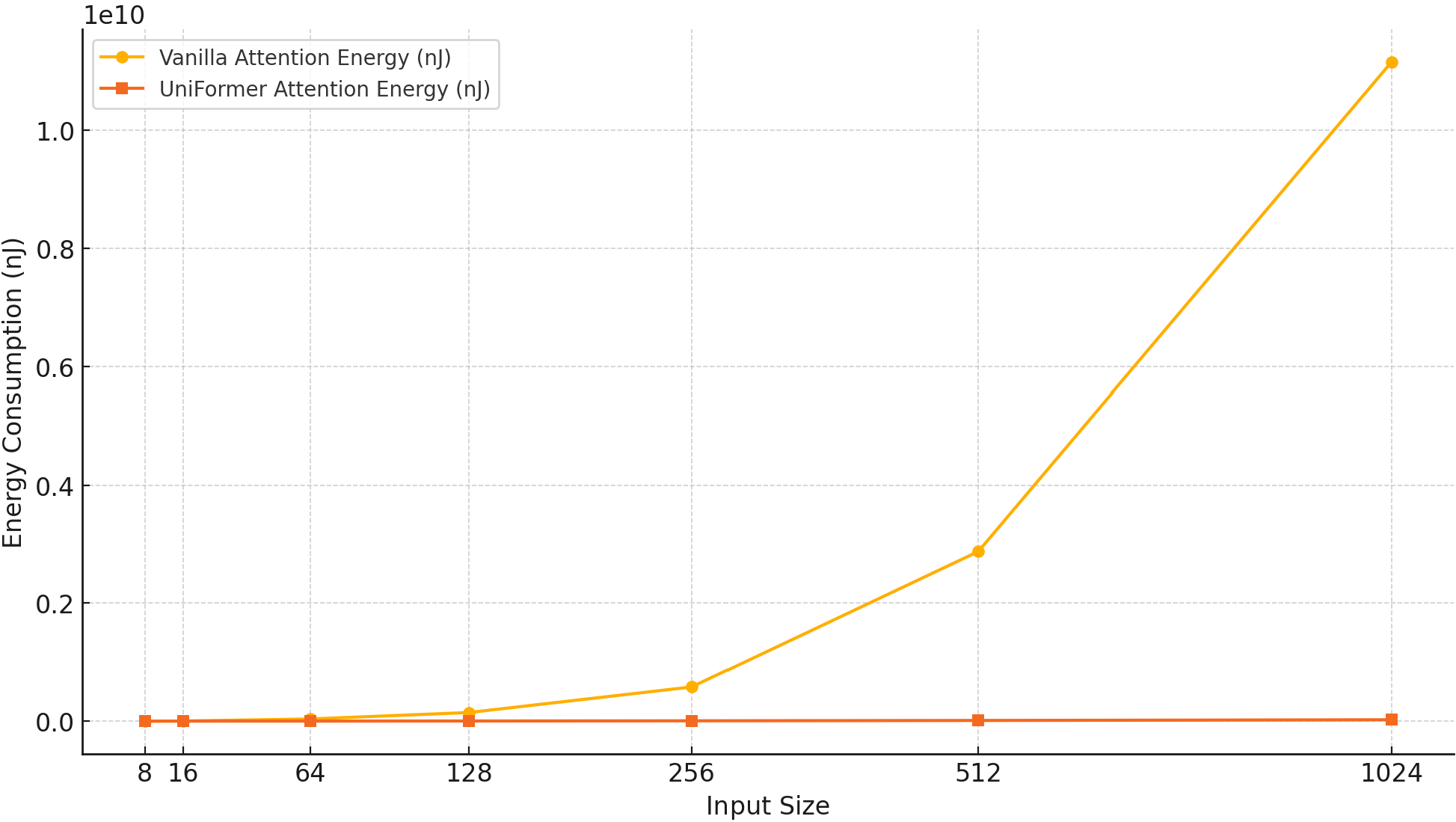}
    \caption{Energy efficiency comparison between Vanilla and UniFormer attention mechanisms.}
    \label{fig:efficiency_comparison}
\end{figure*}

\textbf{Experiment Set Up:} We implement our FPGA design using the AXU15EG development board by Xilinx Vitis HLS 2024.1, which integrates a Xilinx Zynq UltraScale+ MPSoC XCZU15EG-2FFVB1156 chip. A typical operating frequency of 250 MHz is assumed for all experiments. The estimated power consumption is based on a typical AXU15EG power profile of 25W.

\end{document}